# Operator Imprecision and Scaling of Shor's Algorithm

C. Ray Hill and George F. Viamontes*
Lockheed Martin Advanced Technology Laboratories
3 Executive Campus
Cherry Hill, NJ 08002

Shor's algorithm (SA) is a quantum algorithm for factoring integers. Since SA has polynomial complexity while the best classical factoring algorithms are sub-exponential, SA is cited as evidence that quantum computers are more powerful than classical computers. SA is critically dependent on the Quantum Fourier Transform (QFT) and it is known that the QFT is sensitive to errors in the quantum state input to it. In this paper, we show that the polynomial scaling of SA is destroyed by input errors to the QFT part of the algorithm. We also show that Quantum Error Correcting Codes (QECC) are not capable of suppressing errors due to operator imprecision and that propagation of operator precision errors is sufficient to severely degrade the effectiveness of SA.  Additionally we show that operator imprecision in the error correction circuit for the Calderbank-Shor-Steane QECC is mathematically equivalent to decoherence on every physical qubit in a register. We conclude that, because of the effect of operator precision errors, it is likely that physically realizable quantum computers will be capable of factoring integers no more efficiently than classical computers.

**Summary**
It was realized very early in the history of quantum computing that quantum systems are exquisitely sensitive to external disturbances and that management of the resulting computational errors is essential for successful operation of any quantum computer [1, 2]. The development of quantum error correction techniques demonstrated that it is possible to suppress those errors that are most likely to occur due to external disturbances. On the other hand, environmental disturbances are not the only source of errors in a quantum computer, since the operations that drive the computation can never have total precision. While quantum error correction is able to suppress the discrete errors due to environmental disturbances, quantum error correction has no effect on operator precision errors. Because quantum computation is a kind of continuous computation, operator precision errors propagate and grow during the course of a computation. Even relatively small precision errors on each individual operation can result in significant cumulative error at the end of a large computation.

The most famous application of quantum computing is Shor's algorithm for factoring integers [3]. Shor's algorithm has polynomial complexity while the most efficient classical algorithm for factoring integers has sub-exponential complexity. This fact suggests that quantum computation might have some intrinsic advantage over classical computation and might be capable of efficiently solving problems for which no efficient classical solution is possible. The polynomial complexity of Shor's algorithm refers to the number of "quantum gate" operations required to implement it and does not necessarily reflect the scaling of execution time. Because Shor's algorithm, like all quantum computations, yields probabilistic results, the time required for execution depends on the probability of obtaining a desired outcome during a single computation. This fact has important ramifications since the outcome probability of quantum computations is sensitive to errors that accumulate during the computation. For the case of Shor's algorithm, it is possible to prove that operator precision errors severely degrade the performance of the algorithm in such a way that the polynomial complexity of the algorithm no longer describes the scaling behavior of the algorithm's execution time. It is possible to show that, in the presence of operator precision errors, Shor's algorithm is no more efficient than classical algorithms.

---
*Mr. Viamontes now works as a independent consultant.

**Operator Imprecision and Errors in Quantum Computation**
Two sources of errors are important in quantum computation: (1) environmental disturbances of the quantum state of the system that realizes the computation (decoherence), and (2) imprecision of the physical operations that are carried out to implement the computational algorithm [1, 2]. Errors due to environmental disturbances have been the main focus of analysis in the quantum computing literature; however, operator imprecision has a significant impact on quantum algorithms because of error propagation and growth. In addition, it turns out that, while some environmental errors and contrived forms of operator imprecision can be suppressed by quantum error correction codes, the realistic case of small operator imprecision error in every gate is unaffected by error correction. In fact quantum error correction is only able to reliably suppress environmental and operator imprecision errors that affect only one or a few physical qubits in a Quantum error-correcting code (QECC) code block [4]. We will show, however, that if even a small amount of imprecision is present in every gate, then all qubits in every code block will be affected, and more importantly the error in any given qubit propagates to all other qubits.

Quantum computation is a kind of continuous computation. The most familiar example of continuous computation is analog computation, which is central to analog electronics and other analog technologies such as hydraulics and pneumatics. The physical realization of a quantum computation must be a quantum system, i.e., a physical system that must be described by quantum physics rather than classical physics. Quantum physics differs from classical physics in the way that observables are computed and interpreted. While all real-world measurements have dispersion associated with them, measurements of the observables of a quantum system have an irreducible dispersion that is completely independent of any kind of error. This means that measurements of the observables of a quantum system will have some dispersion even though the state of the system is known with total precision. The notion of the state of a system is fundamental to quantum physics. In classical physics, system state does not have a well defined meaning, and the functional values of system observables fill the role in classical physics that the concept of system state performs in quantum physics.

The state of a quantum system is described by a state vector. For the cases of interest to quantum computation, the state vector exists in a $2^L$ dimensional complex vector space where L is the number of qubits. Each qubit is a two state quantum system that has an independent observable with two possible outcomes, usually denoted $|0\rangle$ and $|1\rangle$. This is the traditional representation of state vectors, called the Dirac braket notation. A state vector denoting a basis vector in the entire $2^L$ dimensional state space is written with a binary number, each digit representing the outcome of a measurement on a particular qubit. The evolution rule for a non-relativistic quantum system is given by the Schrödinger equation:

$$i\hbar \frac{\partial |\psi\rangle}{\partial t} = H|\psi\rangle$$

where H is a Hermitian operator called the Hamiltonian, $|\psi\rangle$ is the state vector, and $\hbar$ is Planck's constant. The solution to this equation has the form

$$|\psi(t)\rangle = U(t)|\psi(0)\rangle$$

$$U(t) = e^{-\frac{i}{\hbar}\int_0^t H(\tau)d\tau}$$

The unitary operator U(t) is called the evolution operator. In a typical quantum algorithm, a register of L qubits is initialized (usually to |000…0>), a sequence of evolutions is applied to the qubits in the register, and measurements are performed on each qubit to produce an integer outcome. We can represent this computation with the following equation:

$$|\psi\rangle = U_N...U_2 U_1 |000...0\rangle$$

where $U_i$ is the unitary evolution for the $i^{th}$ "quantum gate" in the network. It is clear from this that each evolution $U_i$ is a continuous computation. Because measurements between evolution steps are not possible for quantum computations, any errors in the outcome of individual evolutions propagate through the computation. It is also clear from this discussion that operator errors can arise from either errors in the physical implementation of the Hamiltonian that drives the operation or from errors in operation time duration. Operator errors are classical in character and are expected to have the classical statistical character of a continuous distribution.

**Quantum Error Correction and General Error Syndromes**
QECCs have been shown to be effective at suppressing one- or a few-qubit errors arising in quantum states of interest to quantum computing. A QECC is a subspace of the complete state space of a quantum system, together with error detection and error correction operators that detect and correct error excursions of code states within the subspace [4]. If C is a quantum code, then any state $|c\rangle \in C$ represents a valid code state. Let the set of operators $\{E_0 = \mathbf{I}, E_1, E_2, ...\}$ be the set of discrete errors that are correctable by the QECC. With the exception of $E_0 = \mathbf{I}$, the action of one of these operators on a state $|c\rangle \in C$ produces a vector $|\psi\rangle \notin C$ that has a component outside of the coding subspace C. The detection operator D on C can be defined by its action on a state $|\psi\rangle$ that has some correctable error, augmented by an ancilla state $|a\rangle$

$$D|\psi\rangle|a\rangle = \sum_e \alpha_e E_e P|\psi\rangle|e\rangle$$

The ancilla state allows detection of errors in the $|\psi\rangle$ state without disturbance of the integrity of the state. The operator P is the projection operator onto the subspace C. Measurement of the ancilla state reveals the specific error that was present in $|\psi\rangle$. There are two results of measuring the ancilla state: (1) the value of this measurement specifies the error that was present in $|\psi\rangle$, and (2) the state of the code becomes $|\psi_d\rangle = kE_e P|\psi\rangle$ which is the result of projection of $|\psi\rangle$ onto the subspace C followed by application of the error $E_e$. Correction of the error is now possible by application of the inverse error operator corresponding to the measured error syndrome

$$|\psi_c\rangle = E_e^{-1}|\psi_d\rangle$$

The QECC scheme described above is not able to correct all possible errors, only those errors that have been included in its construction. The most general error syndrome consists of a superposition of a code state $|c\rangle$ and a random error state $|\varepsilon\rangle$

$$|\psi\rangle = a|c\rangle + b|\varepsilon\rangle$$

The action of the detection operator on this state is

$$D(a|c\rangle + b|\varepsilon\rangle)|a\rangle = \sum_e \alpha_e E_e (a|c\rangle + bP|\varepsilon\rangle)|e\rangle$$

It is clear from this that the projection of the error state $|\varepsilon_c\rangle = P|\varepsilon\rangle$ onto C is not correctable. It is therefore easy to see that application of quantum error correction in the presence of a general error syndrome has limited mitigating effect on errors that have components in C. The uncorrectable component of error is the projection of the error state onto C and this error is unaffected by the error correction code. It might seem that the fact that $|\varepsilon_c\rangle$ is a projection reduces the magnitude of the error, but this is not the case since the number of bits of information carried by the code subspace C is commensurate with the projected error state. If the code subspace C represents *n* bits of information, then the projected error is the error associated with *n* bits.

**Operator Imprecision and Fault Tolerance**
The use of quantum error correcting codes places restrictions on the way that operations can be performed by a quantum computer since only a restricted set of quantum gate operations are feasible on encoded blocks of qubits [4, 5]. This fact is not considered to be a problem since it is possible to approximate arbitrary quantum gate operations as a sequence of "universal" gate operations; however, the presence of operator imprecision errors limits the effectiveness of this strategy.

The Solovay-Kitaev theorem [6] states that an arbitrary unitary transform in SU(2) can be approximated by a sequence of transforms from a universal set of transforms and that the length of the sequence N is related to the approximation precision *p* by

$$N = \log^c\left(\frac{1}{p}\right)$$

where $c \approx 3.97$.

If the unitary transforms in the sequence have random, uncorrelated fidelity errors then the approximate unitary transform formed by the sequence has an overall fidelity error $f$

$$f = f_0 \log^{\frac{c}{2}}\left(\frac{1}{p}\right)$$

This result assumes that the errors in the sequence propagate as a random walk. The overall error in the approximate transform is

$$\varepsilon^2 = f^2 + p^2$$

This expression has a minimum value. This means that a unitary transform approximated by Solovay-Kitaev decomposition has an irreducible minimum error that is determined by the fidelity errors of the constituent transforms of the decomposition.

**Number Theory of Shor's Algorithm**

Shor's Algorithm (SA) is based on the equivalence of factoring integers to finding the period of a modular exponential function [2, 3]

$$a^r = 1 \mod N$$

where $N = PQ$ is an odd composite integer to be factored. This equivalence can be understood by considering the following equations

$$s^2 - 1 = 0 \mod N = mN$$
$$s + 1 \neq 0 \mod N$$
$$s - 1 \neq 0 \mod N$$

where $m$ is an integer. This means that $s^2 - 1$ is a multiple of N but s+1 and s-1 are not.

If we can find a number $s$ that satisfies these conditions, then it is easy to see that either s+1 or s-1 have a common divisor with N

$$(s+1)(s-1) = mN$$

We can find such a number s by identifying s with $a^{\frac{r}{2}}$.

$$\left(a^{\frac{r}{2}} + 1\right)\left(a^{\frac{r}{2}} - 1\right) = mN$$

Thus, if we randomly choose a number $1 < a < N$ and find the period of $a^r \bmod N$, then the following equation defines a factor of N

$$P, Q = \gcd\left(N, a^{\frac{r}{2}} \pm 1\right)$$

It can be shown that there is a high probability of success with a random choice of $a$ [2, 3].

The naïve classical algorithm to find the period $r$ is to simply compute the modular exponential function starting at x=1 and continuing until the result is 1 modN. This is only efficient if the period of the function is known to be small compared to N. If it were possible to always choose a such that, with high probability, the period $r$ is small compared to N, then this simple classical algorithm would be efficient. The reason that this is not possible is the lack of knowledge about the distribution of periods $r$ for different choices of $a$.

Shor's Algorithm uses a quantum computation to find the period of the modular exponential function. This quantum computation is efficient (polynomial in logN) in the absence of errors in the quantum computer carrying out the calculation. When errors are introduced to the QFT period finding part of the quantum computation, the ability of the algorithm to find short periods is degraded. Thus in the presence of errors, Shor's Algorithm is reduced to an analogous situation as the simple classical algorithm: in each case, the probability of finding a period with a fixed amount of computation depends on the distribution of periods with a randomly chosen a. The classical algorithm succeeds for small periods but not for large periods while the reciprocal situation obtains for the quantum algorithm which succeeds for large periods but not for small periods.

A lower bound on the period for a given a can be found starting with the equation defining the periodicity of the modular exponential function

$$a^r - 1 = mN$$

Taking the log of both sides gives

$$r \log(a) = \log(mN + 1)$$

If we ignore the one on the right hand side and rearrange, we get

$$r \approx \frac{\log(m) + \log(N)}{\log(a)}$$

There is no way to calculate m, but the following is the lower bound for r as a function of a

$$r \geq \frac{\log(N)}{\log(a)}$$

This lower bound is not particularly useful; however, the form of the expression suggests that the distribution of r should scale logarithmically with increasing N.

**Effect of Input Errors on the Quantum Fourier Transform**

Shor's algorithm finds the period of modular exponential functions by first creating a quantum state that is a superposition of all values of the function and then applying the Quantum Fourier Transform [3]. Errors that accumulate during the modular exponentiation are input to the QFT and result in degraded performance of period determination. Following Barenco et al. [7], consider the following quantum state as input to the QFT

$$|\Psi\rangle = \frac{1}{\sqrt{N}} \sum_{a=0}^{2^L-1} f(a)|a\rangle$$

where

$$f(a) = \delta_{l, a \bmod r}$$

and

$$N = \left\lceil \frac{2^L}{r} \right\rceil$$

$f(a)$ is a periodic function with period r with offset l, which is smaller than r. The rational for choosing this state is discussed in [7]. The QFT of this state is

$$|\tilde{\Psi}\rangle = \sum_{c=0}^{2^L-1} \tilde{f}(c)|c\rangle$$

and $\tilde{f}(c)$ is non-zero only for values of c that are very close to multiples of $\frac{2^L}{r}$. For the purposes of this analysis, it suffices to assume that $\tilde{f}(c)$ is non-zero at exactly r values of c and all have equal magnitudes at these c values.

Now consider the effect of adding random errors to $f(a)$

$$|\Psi\rangle = \frac{1}{\sqrt{N'}} \sum_{a=0}^{2^L-1} (f(a) + \varepsilon(a))|a\rangle$$

and the QFT becomes

$$|\tilde{\Psi}\rangle = \frac{1}{S} \sum_{c=0}^{2^L-1} \left( \tilde{f}(c) + \tilde{\varepsilon}(c) \right) |c\rangle$$

If we ignore the cross terms between $\tilde{f}(c)$ and $\tilde{\varepsilon}(c)$ then the normalization constant is

$$S^2 = \sum_{c=0}^{2^L-1} \left( \left|\tilde{f}(c)\right|^2 + \left|\tilde{\varepsilon}(c)\right|^2 \right)$$

We will assume that the values $\left|\tilde{f}(c)\right|^2$ are equal and nonzero at r values of c and that the values of $\left|\tilde{\varepsilon}(c)\right|^2$ are equal at all values of c, in which case

$$S^2 = r\left|\tilde{f}\right|^2 + \left(2^L - r\right)\left|\tilde{\varepsilon}\right|^2$$

It is clear from this that the probability of measuring a state $|c\rangle$ where $\tilde{f}(c)$ is non-zero (and thereby successfully measuring the period of the function $f(a)$) is large only if the following condition is true

$$r\left|\tilde{f}\right|^2 \gg \left(2^L - r\right)\left|\tilde{\varepsilon}\right|^2$$

This condition means that the "signal to noise" ratio for the input state to the QFT must be much larger than $\frac{2^L}{r}$

$$\frac{\left|\tilde{f}\right|^2}{\left|\tilde{\varepsilon}\right|^2} \gg \frac{2^L}{r}$$

This condition places a lower bound on the value of the period, r, that can be detected by the QFT in the presence of noise.

**Experimental Results**
To demonstrate the inability of QECC to suppress operator precision errors, we analyze simulation results of a particular 7-qubit register QECC scheme from the class of Calderbank-Shor-Steane (CSS) encodings [5]. This register is flipped by logical NOT gates over 700 times, and each time the CSS error correction scheme is applied to correct any superposition of bit- and phase-flip errors. Decoherence effects are ignored to demonstrate that gate imprecision alone limits the functionality of Shor's algorithm even when a QECC is employed to combat the gate-induced error.

The simulation uses the QuIDDPro quantum circuit simulation package [8] to implement the 7-qubit CSS register. The register is initialized to the logical 0 state using the circuit shown in Figure 1. Seven physical qubits are initialized to the ground state and then put into a superposition of a 7-bit Hamming style codeword

$$|\bar{0}\rangle = \frac{1}{\sqrt{8}}(|0000000\rangle + |0001111\rangle + |0110011\rangle + |0111100\rangle + |1010101\rangle + |1011010\rangle + |1100110\rangle + |1101001\rangle)$$

Flipping any qubit, even by a small probability amplitude, creates a state in the superposition that is not part of the logical 0 codeword. If the amplitude of the flip is high enough, the error may be projected out of the logical 0 state and subsequently corrected.

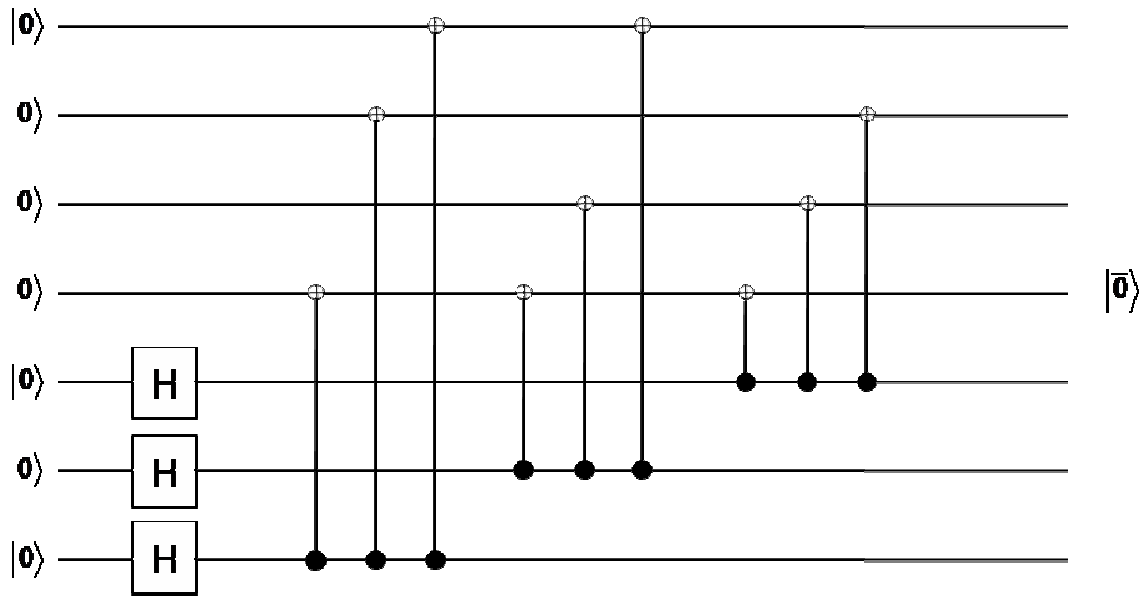

**Figure 1. Initializes seven physical qubits to the logical 0 state in the CSS scheme.**

To detect and correct such errors in the register, three ancilla qubits are used to project bit flip errors as parity changes. The register with the parity checks on the ancilla qubits is shown in Figure 2.

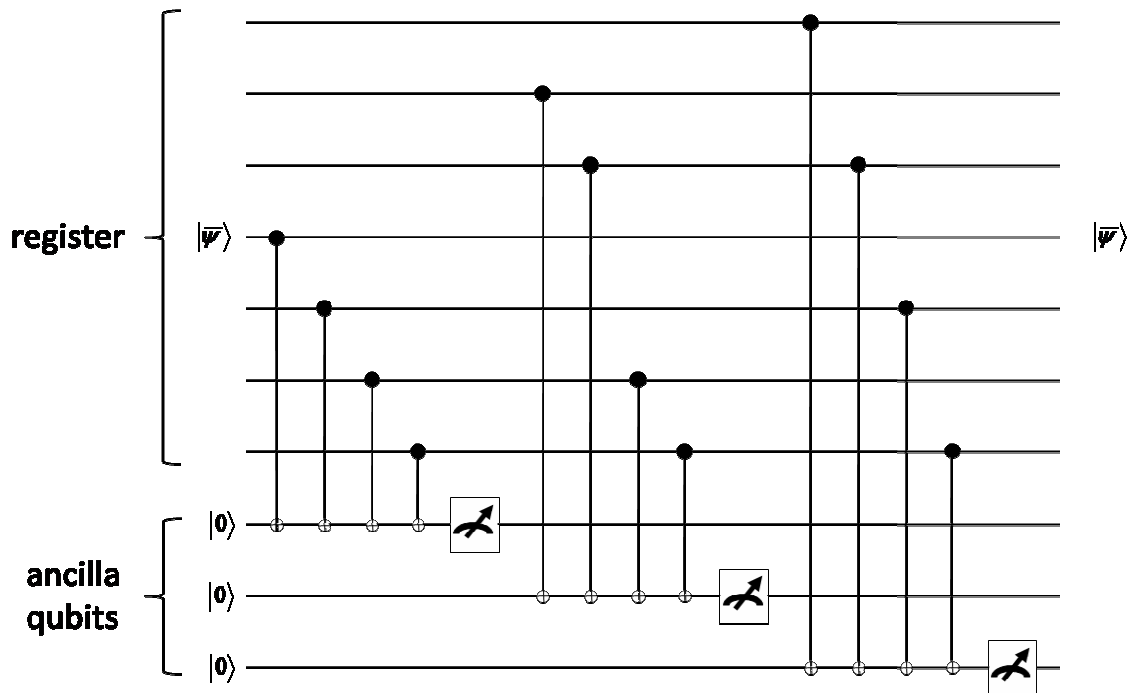
**Figure 2. Circuit that detects bit flip errors in a 7-qubit CSS register.**

The meter symbol denotes measurement of the ancilla qubits. The measured binary values of the ancilla qubits denote the physical qubit wire number that a single bit-flip error occurred on. For example, if the three ancilla qubits are measured as |101> then a bit-flip occurred on the fifth qubit in the register (numbering proceeds from top to bottom in binary). Upon detecting an error, an X gate may be applied to the affected physical qubit wire to flip its state back to the correct state [5]. To detect phase flip errors, the same circuit is used except that Hadamard gates are applied to the register qubits at the beginning and end of the circuit. The effect of the Hadamard gates is to move the register qubits from the computational basis to the phase basis [2]. In phase space, X gates act as phase-flipping rather than bit-flipping operators. Thus, both bit and phase flip errors may be detected by applying the circuit twice, where one application is conjugated by Hadamard gates. In either basis a measurement outcome of |000> denotes no errors detected [5]. Notice that the register may be in any state, and in the case of an error-free circuit, the detection does not change the state of the register. This is a crucial point as it demonstrates that the CSS code, like other QECCs, projects out any errors without revealing information about the state of the register, thus avoiding inadvertent dechoerence of the register's data [5].

Unfortunately the gates shown in Figures 1 and 2 are not fault-free in reality in that they contain some $\varepsilon$ of rotation error. Expanding on the previous discussion of gate imprecision error, virtually all implementations of quantum computers use some form of EM radiation to implement operators, and the type of operator is controlled by the duration and frequency of the EM pulse [2]. The duration and frequency of these operator pulses rotates the affected qubits, and since neither can be controlled perfectly in a real device, over- and under-rotations of the qubits from their ideal target values occur every time an operator is applied. This error is modeled in the following way for the 1-qubit gate set consisting of the Hadamard, NOT, phase,

and identity gates, which are all used in the modular exponentiation simulation with the CSS code

$$H(\varepsilon) = \begin{bmatrix} \cos\left(\frac{\pi}{4}+\varepsilon\right)i & \sin\left(\frac{\pi}{4}+\varepsilon\right)i \\ \sin\left(\frac{\pi}{4}+\varepsilon\right)i & -\cos\left(\frac{\pi}{4}+\varepsilon\right)i \end{bmatrix}, \quad X(\varepsilon) = \begin{bmatrix} \cos\left(\frac{\pi}{2}+\varepsilon\right) & \sin\left(\frac{\pi}{2}+\varepsilon\right) \\ \sin\left(\frac{\pi}{2}+\varepsilon\right) & -\cos\left(\frac{\pi}{2}+\varepsilon\right) \end{bmatrix}$$

$$Z(\varepsilon) = \begin{bmatrix} \cos(\varepsilon) & \sin(\varepsilon) \\ \sin(\varepsilon) & -\cos(\varepsilon) \end{bmatrix}, \quad I(\varepsilon) = \begin{bmatrix} \cos(\varepsilon) & -\sin(\varepsilon) \\ \sin(\varepsilon) & \cos(\varepsilon) \end{bmatrix}$$

The key 2-qubit used in the above circuits is the CNOT gate. The error modeling for this gate is similar except that each control and target portion of the operator representation makes use of potentially different rotation errors

$$CNOT(\varepsilon_0,\varepsilon_1) = \begin{bmatrix} \cos(\varepsilon_0) & -\sin(\varepsilon_0) & 0 & 0 \\ \sin(\varepsilon_0) & \cos(\varepsilon_0) & 0 & 0 \\ 0 & 0 & \cos\left(\frac{\pi}{2}+\varepsilon_1\right) & \sin\left(\frac{\pi}{2}+\varepsilon_1\right) \\ 0 & 0 & \sin\left(\frac{\pi}{2}+\varepsilon_1\right) & -\cos\left(\frac{\pi}{2}+\varepsilon_1\right) \end{bmatrix}$$

In the CSS scheme, any of the above gates can be applied to the logical 7-qubit register simply by applying the gate piece-wise to each physical qubit. We now show that by attempting to correct imprecision errors caused by these faulty gates, the error detection and correction circuit in Figure 2 not only introduces more gate imprecision error, but the operators involving the ancilla qubits are mathematically equivalent to phase damping decoherence. To demonstrate this equivalence, consider the circuit representation for phase damping decoherence

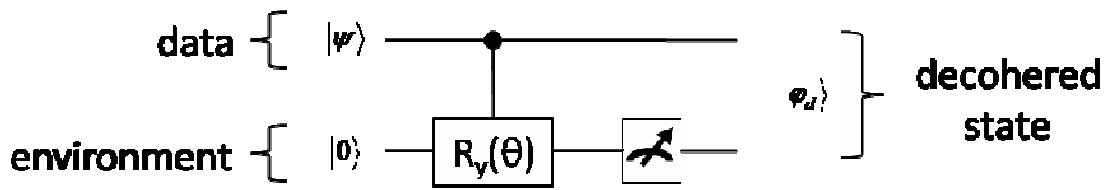

**Figure 3. Phase damping decoherence model for the environment acting on a qubit.**

Note that this decoherence model very closely resembles any of the CNOT gates of Figure 2 in which a physical qubit from the register is the control, an ancilla qubit initialized to the |0> state is the target, and the same ancilla qubit is measured. In Figure 3 the very same roles are played by the data qubit and environment qubit, respectively. In fact the only difference between Figure 3 and any CNOT gate in Figure 2 is that the decoherence model makes use of a controlled-$R_y(\theta)$ gate rather than a faulty CNOT gate. A fault-free CNOT gate is the desired behavior in Figure 2, so strictly speaking the CNOT gate itself is not a source of unwanted decoherence. However, since the CNOT gate is faulty, it can be shown that ε over- or under-rotation component is indeed equivalent to the decoherence model shown in Figure 2. To demonstrate this mathematically, consider decomposing a faulty CNOT gate as follows

$$CNOT'(\varepsilon_0, \varepsilon_1) = \begin{bmatrix} \cos(\varepsilon_0) & -\sin(\varepsilon_0) & 0 & 0 \\ \sin(\varepsilon_0) & \cos(\varepsilon_0) & 0 & 0 \\ 0 & 0 & 0 & 1 \\ 0 & 0 & 1 & 0 \end{bmatrix} \begin{bmatrix} 1 & 0 & 0 & 0 \\ 0 & 1 & 0 & 0 \\ 0 & 0 & \cos(\varepsilon_1) & \sin(\varepsilon_1) \\ 0 & 0 & -\sin(\varepsilon_1) & \cos(\varepsilon_1) \end{bmatrix}$$

By multiplying the two matrices together and applying trigonometric identities, the reader can verify that this decomposition is equivalent to the previous expression given for the faulty CNOT gate. The matrix on the left is equivalent to a CNOT with $\varepsilon_0$ rotation error on the control qubit portion. The matrix on the right is similar to an identity matrix with $\varepsilon_1$ rotation error on the target qubit portion. However, the matrix on the right is in fact equivalent to the controlled $R_y$ gate in Figure 3 up to a single X gate conjugation.

To prove this, consider the matrix for a controlled $R_y$ gate

$$R_y = \begin{bmatrix} 1 & 0 & 0 & 0 \\ 0 & 1 & 0 & 0 \\ 0 & 0 & \cos(\theta/2) & -\sin(\theta/2) \\ 0 & 0 & \sin(\theta/2) & \cos(\theta/2) \end{bmatrix}$$

If we let U denote the matrix on the right of the faulty CNOT decomposition and V denote the matrix for the controlled $R_y$ gate, then both matrices are related as follows

$$V = (I \otimes X)U(I \otimes X)$$

In other words, V is equivalent to U with the target qubit conjugated by a simple X gate. As shown in Figure 3, applying U followed by measurement on the target qubit is a model of decoherence. The fautly CNOT gates in Figure 2 are followed by measurement on the target (ancilla) qubits. Also, the first extraneous X gate may be viewed as part of a computation that precedes the decoherence effect, while the second extraneous X gate can be merged with the projective measurement operator. The decomposition therefore proves that gate imprecision error in the CSS error correction circuit is equivalent to decoherence. As shown in Figure 2, this decoherence model is effectively applied to each physical qubit in the register. QECCs, including CSS, cannot correct multiple-qubit errors of non-trivial magnitudes [4, 5], which presents a significant problem for the CSS scheme.

To estimate an optimistic bound for the robustness provided by CSS in Shor's algorithm in light of this problem, we performed a simulation that applies over 700 logical NOT operators to the 7-qubit register. The logical NOTs are implemented by applying a faulty X gate to each physical qubit individually. The purpose of applying this operation is to get an estimate of what would happen to a logical qubit register during the modular exponentiation phase of Shor's algorithm [3]. All known implementations of this circuit, including the arithmetic circuit described by Vedral et al. [9], are equivalent to circuits that make heavy use of controlled NOT operations between logical qubits. In Vedral et al.'s circuit implementation, a 100 logical qubit instance of

Shor's algorithm requires approximately 100,000 applications of a reversible ripple-carry adder, where each adder circuit will apply half a dozen controlled NOT and multiply controlled NOT gates to each logical qubit [9]. Since only a single register is implemented, very little entanglement is generated in the circuit. Coupling this with the fact that just over 700 operators are applied to the register, this simulation provides an optimistic bound on the error rate. The simulation data is shown in Figure 4.

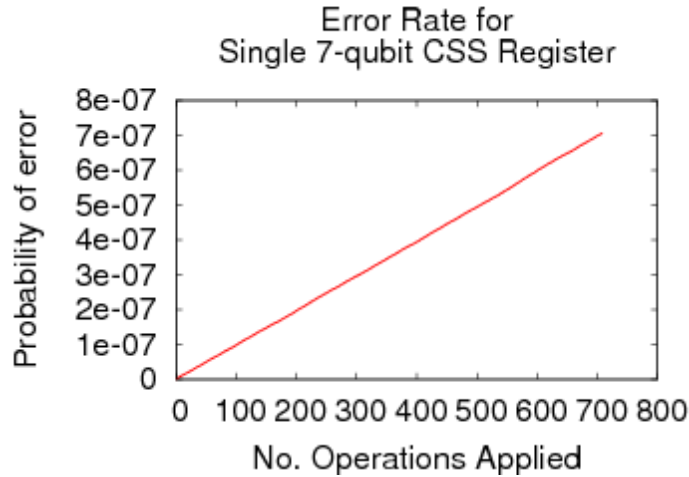

**Figure 4. Error data for a single 7-qubit CSS register with a +-1e-5 rotation error range.**

Although the final probability of error after over 700 NOTs are applied is quite small, the data indicates a linear trend with respect to the number of operations. The observed linear growth of error with number of operations is consistent with propagation of operator precision errors as a random walk process. As noted above, 100 logical qubits, or 700 physical qubits, would require well over 100,000 operations. This would be sufficient to factor a 40-50 bit number given an ideal physical device [2]. However, since 256-bit RSA is fairly common today, Shor's algorithm with the CSS ECC would require well over 500 logical qubits, or 3,500 physical qubits. Modular exponentiation for such a circuit size would require well over a quarter of a million operators. Given that the simulation data represents an optimistic upper bound for error resilience in the face of gate imprecision, robust quantum integer factoring with Shor's algorithm seems to have a hard upper limit even for a decoherence-free device.

**Concluding Remarks**
Quantum error correction is capable of reliably suppressing single-qubit errors due to environmental disturbances or operator precision errors. However, operator imprecision errors propagate and grow during the course of a quantum computation and have an important impact on the efficiency of the computation. In particular, we have shown that operator precision errors break the polynomial scaling of Shor's algorithm and conclude that, in the presence of operator precision errors, Shor's algorithm is no more efficient than classical algorithms for factoring integers. To demonstrate how operator precision errors propagate in practice, we proved that the error correction circuit for the CSS QECC is mathematically equivalent to applying decoherence on *each* physical qubit in a logical qubit register. We then used simulation to show that this accumulated error on each qubit causes the probability of error in a CSS QECC register to increase linearly with the number of gates applied.

It is natural to ask whether these results have wider implications about the power of quantum computers relative to classical computers. While the results presented in this paper do not answer this question definitively, it is important to note the singular stature of Shor's algorithm as the only quantum algorithm that appears to efficiently solve a classically intractable problem. The fact that Shor's algorithm is not more efficient than classical algorithms removes the only strong evidence for the superior computational power of quantum computers relative to classical computers. Expanding on the theoretical results presented in this work and applying other classical simulation algorithms [10, 11, 12, 13] to other QECC schemes can significantly strengthen this argument.